\documentclass[prb,showpacs,floatfix,twocolumn,eqsecnum]{revtex4-1}
\usepackage{graphicx}
\usepackage{multirow}
\usepackage{dcolumn}
\usepackage{bm}
\usepackage[normalem]{ulem}
\usepackage{amsmath}
\usepackage{amsfonts}
\usepackage{amssymb}
\usepackage{color}
\usepackage{colordvi}

\begin{document}

\title{Quantum criticality in two dimensions and Marginal Fermi Liquid}
\date{\today }
\author{K.B. Efetov$^{1,2,3}$}
\affiliation{$^1$ Institut f\"ur Theoretische Physik III, Ruhr-Universit\"at Bochum,
44780 Bochum, Germany\\
$^2$ National University of Science and Technology ``MISiS'', Moscow,
119049, Russia\\
$^3$ International Institute of Physics, UFRN, 59078-400 Natal, Brazil}

\begin{abstract}
Kinetic properties of a two dimensional model of fermions interacting with
antiferromagnetic spin excitations near the quantum critical point (QCP) are
considered. The temperature or doping are assumed to be sufficiently high,
such that the pseudogap does not appear. In contrast to standard
spin-fermion models, it is assumed that there are intrinsic inhomogeneities
in the system suppressing space correlations of the antiferromagnetic
excitations. It is argued that the inhomogeneities in the spin excitations
in the \textquotedblleft strange metal\textquotedblright\ phase can be a
consequence of existence of \textquotedblleft $\pi $-shifted%
\textquotedblright\ domain walls in the doped antiferromagnetic phase.
Averaging over the inhomogeneities and calculating physical quantities like
resistivity and some others one can explain unusual properties of cuprates
unified under the name \textquotedblleft Marginal Fermi
Liquid\textquotedblright\ (MFL). The dependence of the slope of the linear
temperature dependence of the resistivity on doping is compared with
experimental data.
\end{abstract}

\maketitle

\section{Introduction}

Properties of the normal state\ of high $T_{c}$ superconducting cuprates in
the vicinity of the quantum critical point (QCP) are not consistent with the
Landau Fermi liquid theory. Such unusual effects as the linear dependence of
the resistivity on temperature, the linear tunnelling conductivity as a
function of voltage, almost frequency and temperature independent
backgrounds in the Raman-scattering intensity, constant thermal
conductivity, and a very large nuclear relaxation time are similar in all $%
CuO$ based high-$T_{c}$ compounds. This region is usually referred to as
\textquotedblleft strange metal\textquotedblright .

In the pioneering work Varma et al \cite{varma1,abrahams0} have proposed a
\textquotedblleft marginal Fermi liquid\textquotedblright\ (MFL)
phenomenology that allowed them to describe the unusual experimental
findings surprisingly well. The theory is based on the assumptions that 1)
electrons are scattered by unknown bosonic excitations characterized by a
retarded propagator $\chi ^{R}\left( \mathbf{q,}\omega ,T\right) ,$ where $%
\mathbf{q}$ is momentum, $\omega $ is frequency and $T$ is temperature, 2)
the imaginary part of this propagator has the form%
\begin{equation}
\mathrm{Im}\chi ^{R}\left( \mathbf{q,}\omega ,T\right) =\left\{
\begin{array}{cc}
\nu \left( \omega /T\right) , & \omega \ll T \\
\nu \left( sgn\omega \right) , & T\ll \omega \ll \omega _{c}%
\end{array}%
\right. ,  \label{a1}
\end{equation}%
where $\nu $ is the density of states per unit volume and per spin
direction, and $\omega _{c}$ is a high energy cutoff.

Later, Abrahams and Varma \cite{abrahams} have demonstrated that the
marginal Fermi liquid (MFL) assumption described results of angle-resolved
photoemission (ARPES) \cite{kaminski,valla} very well, too (see, also Ref. %
\onlinecite{zhu}).

In spite of the evident success in describing the experiments \cite%
{gurvitch,iye,ando,albenque,cooper}, the final agreement on the origin of
the bosonic mode specified by Eq. (\ref{a1}) seems to be lacking so far. The
strange metal behavior is attributed to quite different phenomena like,
e.g., existence of spontaneous orbital currents \cite{varma2}, quantum
criticality near antiferromagnetic transition \cite{pines,rice,sachdev} and
many others.

A recent discovery of the charge modulation in cuprates \cite%
{julien,ghiringhelli,chang,achkar,sebastian,leboeuf,blackburn,cominSTM}
signals a competition between the superconductivity and a charge density
wave (CDW) in the pseudogap region of the phase diagram of cuprates. Many
important experimental findings of these works can be explained \cite%
{emp,sachdev13a,mepe,hayward,wang} in the framework of the so-called
spin-fermion (SF) model introduced earlier \cite{ac,acs} for description of
electron-electron interaction in the vicinity of QCP. In particular, it has
been proposed in Ref. \onlinecite{emp} that the pseudogap (PG) state arises
as a consequence of the competition between the superconducting and a charge
modulated state.

Experimentally, increasing the temperature and doping one passes from the
pseudogap state to a strange metal state described by the MFL phenomenology.
Assuming that the pseudogap state can be understood in terms of the SF\
model it is natural to use this model also for description of the
\textquotedblleft neighboring\textquotedblright\ strange metal state.
However, the correlation function of antiferromagnetic spin fluctuations
used in the SF model is definitely different from the one given by Eq. (\ref%
{a1}), and new ideas are necessary to overcome this inconsistency.

In this paper we show that the MFL with the bosonic mode, Eq. (\ref{a1}),
can nevertheless be derived from the SF model for the antiferromagnet-normal
metal quantum phase transition in 2D. However, in order to achieve this goal
one should introduce into the model a disorder reducing the
antiferromagnetic correlations at large distances. It is argued that such a
disorder is intrinsically present due to doping and, being sufficiently
smooth, does not contribute to the residual resistivity.

\section{Formulation of the model}

\begin{figure}[h]
\vspace{-0.5cm} \includegraphics[height=1.1in,width=0.5%
\linewidth]{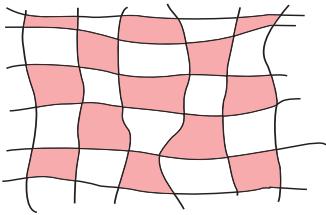} \vspace{0.5cm}
\caption{(Color online.) Domains separated by $\protect\pi $-shifted domain
walls.}
\label{fig01}
\end{figure}

Following this idea we assume that the $CuO$ plains consist of domains $%
\mathrm{f}$, such that the antiferromagnetic (AF) field $\vec{\phi}_{\mathrm{%
f}}$ varies almost periodically with the modulation vector $\mathbf{Q=}%
\left( \pi /b,\pi /b\right) $ inside the domains but sharply changes the
sign when crossing the boarder between them. In other words, the fluctuating
field $\vec{\phi}_{\mathrm{f}}$ is shifted on the boarder by the lattice
period $b$ (the phase of the oscillations is shifted by $\pi $) and we write
it as
\begin{equation}
\vec{\phi}_{\mathrm{f}}\left( \mathbf{r}\right) =I_{\mathrm{f}}\left(
\mathbf{r}\right) \vec{\phi}\left( \mathbf{r}\right) ,\quad I_{\mathrm{f}%
}\left( \mathbf{r}\right) =\left\{
\begin{array}{cc}
1, & \text{\textrm{f }}\in \text{\textrm{\ \textquotedblleft
pink\textquotedblright }} \\
-1, & \text{\textrm{f }}\in \text{\textrm{\textquotedblleft
white\textquotedblright }}%
\end{array}%
\right. ,  \label{a0}
\end{equation}%
In Eq. (\ref{a0}) the field $\vec{\phi}$ is almost periodic everywhere in
space, and \textquotedblleft pink\textquotedblright\ and \textquotedblleft
white\textquotedblright\ domains are represented in Fig. \ref{fig01}.

The size and the form of the domains is not critical at the transition
between the antiferromagnet and paramagnet, and Eq. (\ref{a0}) is assumed to
be applicable on both sides of it. We write the Lagrangian $L$ of the model
as%
\begin{equation}
L=L_{0}+L_{\psi }+L_{\phi }+L_{b}  \label{b1}
\end{equation}

In Eq. (\ref{b1}), $L_{0}$ stands for the Lagrangian of non-interacting
fermions (holes)%
\begin{equation}
L_{0}=\int \psi ^{\ast }\left( \tau ,\mathbf{r}\right) \left[ \partial
_{\tau }+\varepsilon \left( -i\nabla _{\mathbf{r}}\right) -\mu \right] \psi
\left( \tau ,\mathbf{r}\right) d\mathbf{r},  \label{b2}
\end{equation}%
while
\begin{equation}
L_{\psi \phi }=\lambda \sum_{\mathrm{f}}\int \psi ^{\ast }\left( \tau ,%
\mathbf{r}\right) \vec{\sigma}\vec{\phi}_{\mathrm{f}}\left( \tau ,\mathbf{r}%
\right) \psi \left( \tau ,\mathbf{r}\right) d\mathbf{r}  \label{b2a}
\end{equation}%
describes interaction of the fermions with the effective exchange field $%
\vec{\phi}_{\mathrm{f}}\left( \tau ,\mathbf{r}\right) $ of the
antiferromagnet. In Eqs. (\ref{b2},\ref{b2a}), $\psi $ is the anticommuting
fermionic field, $\vec{\sigma}$ is the vector of Pauli matrices, and $\tau $
is the imaginary time. The second term in Eq. (\ref{b2}) stands for the
electron energy operator, and $\mu $ is the chemical potential.

The Lagrangian of $L_{\phi }$ for the exchange field $\vec{\phi}$ is written
near QCP as
\begin{equation}
L_{\phi }=\frac{1}{2}\int \Big[\vec{\phi}\left( \tau ,\mathbf{r}\right) %
\left[ \hat{D}_{0}^{-1}+\frac{g\vec{\phi}^{2}\left( \tau ,\mathbf{r}\right)
}{2}\right] \vec{\phi}\left( \tau ,\mathbf{r}\right) \Big]d\mathbf{r},
\label{b3}
\end{equation}%
where the Fourier transform of $\hat{D}_{0}$ has the form
\begin{equation}
D_{0}\left( \omega _{n},\mathbf{q}\right) =\left( v_{s}^{-2}\omega
_{n}^{2}+\left( \mathbf{Q-q}\right) ^{2}+a\right) ^{-1},  \label{c6}
\end{equation}%
and $\omega _{n}$ is the bosonic Matsubara frequency.

In Eq. (\ref{c6}), $v_{s}$ is the velocity of the spin waves, $a$
characterizes the distance from QCP ($a>0$ on the metallic side and $a<0$ in
the AF region).

Actually, domain walls (DW) separating domains with opposite directions of
the staggered magnetization have been found in 2D using a Hartree-Fock
approximation for a $CuO$ lattice \cite{zaanen} and for the $t$-$J$ model
\cite{poilblanc}, as well as using a mean field approximation for the
Hubbard model \cite{machida}. Similar DW (stripes) have been obtained later
within the $t$-$J$ model numerically using the Density Matrix
Renormalization Group (DMRG) method \cite{white}.

The DW derived in these works separate regions with opposite direction of
the AF ordering ($\pi $-shifted DW). They contain chains of holes in the
middle of DW, while the magnetization vanishes there. According to this
picture, the doped holes are not distributed homogeneously in the AF but are
located inside the DW implying that the doped AF is \emph{intrinsically}
inhomogeneous. The typical distance between the DW is proportional to $%
p^{-1} $ \cite{zaanen,poilblanc,machida,white,tranquada}, where $p$ is the
number of doped holes per $Cu$ atom.

A stripe correlation of spins and holes is evident in cuprates from neutron
diffraction \cite{tranquada,kivelson}. As the DW contain holes, their shape
and locations are affected also by an inhomogeneous electrostatic field of
doping ions located outside the $CuO$ planes. This interaction should make
the shape and size of the domains rather irregular and we assume that Fig. %
\ref{fig01} together with Eqs. (\ref{a0}-\ref{b3}) can properly describe the
antiferromagnet doped with holes.

On the metallic side, $a>0$, field $\vec{\phi}$ can be finite only as a
result of fluctuations. Although the AF order parameter $I_{\mathrm{f}}\vec{%
\phi}_{0}$ vanishes at QCP, the distance between DW determined by the hole
density remains finite at $a=0$.

In the limit of a weak doping $p\sim 0.1-0.2$, the typical size of the
domains $Q_{D}^{-1}\sim \left( Qp\right) ^{-1}$ is considerably larger than
the atomic length $Q^{-1}$, while the length $l_{T}=v_{s}/T$ can be even
larger than $Q_{D}^{-1}$ for relevant temperatures.

Neglecting the quartic term in $L_{\phi },$ Eq. (\ref{b3}), we integrate out
the field $\vec{\phi}$ and come with help of Eq. (\ref{a0}) to action $%
S_{eff}\left[ \psi \right] $
\begin{equation}
S_{eff}\left[ \psi \right] =\int_{0}^{\beta }L_{0}\left[ \psi \right] d\tau
+S_{int}\left[ \psi \right] ,  \label{c3}
\end{equation}%
where
\begin{eqnarray}
&&S_{int}\left[ \psi \right] =-\frac{\lambda ^{2}}{2}\sum_{\mathrm{f,f}%
^{\prime },k=x,y,z}\int d\mathbf{r}d\mathbf{r}^{\prime }d\tau d\tau ^{\prime
}\psi ^{\ast }\left( \tau ,\mathbf{r}\right) \sigma ^{k}\psi \left( \tau ,%
\mathbf{r}\right)  \notag \\
&&\times I_{\mathrm{f}}\left( \mathbf{r}\right) I_{\mathrm{f}^{\prime
}}\left( \mathbf{r}^{\prime }\right) D_{0}\left( \tau -\tau ^{\prime },%
\mathbf{r}-\mathbf{r}^{\prime }\right) \psi ^{\ast }\left( \tau ^{\prime },%
\mathbf{r}^{\prime }\right) \sigma ^{k}\psi \left( \tau ^{\prime },\mathbf{r}%
^{\prime }\right) .  \notag \\
&&  \label{c5}
\end{eqnarray}%
As the DW can randomly be distorted by the potential of the $O$ atoms
located outside the $CuO$ planes, averaging over random $I_{\mathrm{f}%
}\left( \mathbf{r}\right) $ looks a reasonable method of calculation. The
propagator $D_{0}$ varies on distances of order $l_{T}$ and, in the limit $%
l_{T}Q_{D}\gg 1$, one can simply replace the product $I_{\mathrm{f}}I_{%
\mathrm{f}^{\prime }}$ in Eq. (\ref{c5}) by its average. We assume that the
correlations are gaussian with the following moments
\begin{equation}
\left\langle I\left( \mathbf{r}\right) \right\rangle =0,\qquad \left\langle
I\left( \mathbf{r}\right) I\left( \mathbf{r}^{\prime }\right) \right\rangle
=U\left( Q_{D}\left\vert \mathbf{r-r}^{\prime }\right\vert \right) ,
\label{c12}
\end{equation}%
where the function $U\left( x\right) $ decays sufficiently fast at $%
x\rightarrow \infty $ and $U\left( 0\right) =1.$ Eqs. (\ref{b3}-\ref{c12})
fully specify the model considered and allow one to calculate physical
quantities explicitly.

\section{Effective mode.}

Averaging in Eq. (\ref{c5}) over $I\left( \mathbf{r}\right) $ we immediately
come to an effective fermion-fermion interaction $\lambda ^{2}\bar{D}%
_{0}\left( \tau -\tau ^{\prime },\mathbf{r}-\mathbf{r}^{\prime }\right) $
with the propagator%
\begin{equation}
\bar{D}_{0}\left( \tau -\tau ^{\prime },\mathbf{r}-\mathbf{r}^{\prime
}\right) =U\left( Q_{D}\left\vert \mathbf{r-r}^{\prime }\right\vert \right)
D_{0}\left( \tau -\tau ^{\prime },\mathbf{r}-\mathbf{r}^{\prime }\right) .
\label{k1}
\end{equation}%
Eq. (\ref{k1}) shows that the presence of the $\pi $-shifted DW destroys the
spin correlations at distances exceeding the typical domain size $Q_{D}^{-1}$%
.

In the homogeneous case, the bare propagator $D_{0}$ is modified due to the
Landau damping \cite{hertz}. This effect can be obtained in the random phase
approximation (RPA). The polarization function $\Pi \left( \omega _{n},%
\mathbf{q}\right) $ does not depend on $\mathbf{q}$ and is short ranged in
the real space. The function $\Pi \left( \omega _{n},\mathbf{r-r}^{\prime
}\right) $ is essentially non-zero only when both $\mathbf{r}$ and $\mathbf{r%
}^{\prime }$ are located in the same domain. Then, as in the homogeneous
case, one comes to the following relation
\begin{equation}
D^{-1}\left( \omega _{n},\mathbf{q}\right) =D_{0}^{-1}\left( \omega _{n},%
\mathbf{q}\right) -\Pi \left( \omega _{n},\mathbf{q}\right) ,  \label{c7}
\end{equation}%
where%
\begin{equation}
\Pi \left( \omega _{n},\mathbf{q}\right) =C+\gamma \left\vert \omega
_{n}\right\vert ,\quad \gamma =\frac{4\lambda ^{2}}{\pi v^{2}\sin \delta }
\label{c9}
\end{equation}%
and $C$ is a constant renormalizing the position of the QCP. In Eq. (\ref{c9}%
), $v$ is the Fermi velocity at the hotspots and $\delta $ is the angle
between the velocities of the neighboring hot spots (see, e.g., Refs.%
\onlinecite{ac,acs}, and SI of Ref. \onlinecite{emp}). As usual \cite%
{ac,acs,emp}, we neglect the $\omega ^{2}$ -term in the propagator $D\,$,
Eqs. (\ref{c6}, \ref{c7}, \ref{c9}).

Formally, the parameter $a$ entering the propagator $D\left( \omega _{n},%
\mathbf{q}\right) $, Eqs. (\ref{c6}, \ref{c7}), should vanish at the
transition point. However, the transition is smeared in 2D at any finite
temperature by thermal fluctuations. One can estimate the characteristic
width of the transition considering corrections to the coupling constant $g$
within the perturbation theory and keeping only the most divergent static
contributions (SI of \cite{emp}). This gives in the first order
\begin{equation}
g\rightarrow g-Tg^{2}\int \frac{d^{2}k}{\left( k^{2}+a\right) ^{2}},
\label{c20}
\end{equation}%
which leads in the limit $a\rightarrow 0$ to a divergency. Since the
transition is smeared, we conclude that $a$ cannot be effectively smaller
than some minimal value $a_{0}\left( T\right) $ at which the correction in
Eq. (\ref{c20}) is of the same order as the bare coupling $g$. This gives an
estimate for $a_{0}\left( T\right) $
\begin{equation}
a_{0}\left( T\right) =cgT,  \label{c21}
\end{equation}%
where $c$ is a numerical coefficient.

Then, one should replace parameter $a$ in Eq. (\ref{c6}) by
\begin{equation}
a\left( T\right) =a_{0}\left( T\right) +\tilde{a},  \label{c23}
\end{equation}%
where $\tilde{a}$ characterizes the distance from the critical line, to
obtain
\begin{equation}
D\left( \omega _{n},\mathbf{q}\right) =\left( \gamma \left\vert \omega
_{n}\right\vert +\left( \mathbf{Q-q}\right) ^{2}+a\left( T\right) \right)
^{-1}  \label{c11}
\end{equation}

Replacing the function $D_{0}$ in Eq. (\ref{k1}) by $D$ from Eq. (\ref{c11})
one obtains an effective propagator $\bar{D}$ instead of $\bar{D}_{0}$
\begin{equation}
\bar{D}\left( \omega _{n},\mathbf{q}\right) =Q_{D}^{-2}\int \tilde{U}\left(
\frac{\left\vert \mathbf{q-k}\right\vert }{Q_{D}}\right) D\left( \omega _{n},%
\mathbf{k}\right) \frac{d\mathbf{k}}{\left( 2\pi \right) ^{2}},  \label{k2}
\end{equation}%
where $\tilde{U}$ is the Fourier transform of $U$.

The integration over $\mathbf{k}$ makes the propagator $\bar{D}$ weakly
dependent on $\mathbf{q}$ for $\left\vert \mathbf{Q-q}\right\vert \lesssim
Q_{D}.$ The analytical continuation of the propagator $D\left( \omega _{n},%
\mathbf{q}\right) $ from positive Matsubara frequencies $\omega _{n}$ to the
real axis, $i\omega _{n}\rightarrow \omega +i0$, gives the retarded
propagator $D^{R}\left( \omega ,\mathbf{q},T\right) $ that can be obtained
from $D\left( \omega _{n},\mathbf{q}\right) $ by the replacement $\left\vert
\omega _{n}\right\vert \rightarrow i\omega $. Substituting $D^{R}\left(
\omega ,\mathbf{k},T\right) $ instead of $D\left( \omega _{n},\mathbf{k,}%
T\right) $ in Eq. (\ref{k2}) one can obtain the propagator $D^{R}\left(
\omega ,\mathbf{k,}T\right) .$ The real part of $D^{R}\left( \omega ,\mathbf{%
k,}T\right) $ is not interesting for electron transport properties.
Calculation of the integral over two-dimensional momenta $\mathbf{k}$ in Eq.
(\ref{k2}) is performed assuming that the inequality $\gamma \left\vert
\omega \right\vert \lesssim Q_{D}^{2}$ is fulfilled. In this limit, the main
contribution comes from $\left( \mathbf{k-Q}\right) ^{2}\sim \gamma
\left\vert \omega \right\vert \lesssim Q_{D}^{2}$ and the variable $\mathbf{k%
}$ in the function $\tilde{U}$ can be simply replaced by $\mathbf{Q}$. Then,
a straightforward integration over $\mathbf{k}$ (for details, see
Supplementary Information (SI)) provides
\begin{equation}
\mathrm{Im}\bar{D}^{R}\left( \omega ,\mathbf{q},T\right) =\frac{1}{4\pi
Q_{D}^{2}}\tilde{U}\left( \frac{\left\vert \mathbf{q-Q}\right\vert }{Q_{D}}%
\right) \arctan \left( \frac{\gamma \omega }{a\left( T\right) }\right) .
\label{k3}
\end{equation}%
Eq. (\ref{k3}) is in accord with the hypothesis of MFL, Eq. (\ref{a1}), for
temperatures exceeding the distance from the critical line, when $%
a_{0}\left( T\right) \gtrsim \tilde{a}$. Provided this inequality is
fulfilled, and $g$ and $\gamma $ are of the same order (as they should) one
obtains the asymptotics of Eq. (\ref{a1}) in the limits of high $\omega
\gtrsim T$ and low $\omega \lesssim T$ frequencies. The temperature $T$
should also be higher than the coupling energy between the layers, which
guarantees that the spin fluctuations are effectively two-dimensional.

The function $\mathrm{Im}\bar{D}^{R}\left( \omega ,\mathbf{q}\right) $, Eq. (%
\ref{k3}), is generally momentum dependent and thus differs from $\mathrm{Im}%
\chi ^{R}\left( \mathbf{q,}\omega \right) $, Eq. (\ref{a1}). At the same
time, the dependence of $\mathrm{Im}\bar{D}^{R}\left( \omega ,\mathbf{q}%
\right) $, Eq. (\ref{k3}), is rather weak for a small size $Q_{D}^{-1}$ of
the domains and the difference between the functions $\mathrm{Im}\bar{D}%
^{R}\left( \omega ,\mathbf{q}\right) $ and $\mathrm{Im}\chi ^{R}\left(
\mathbf{q,}\omega \right) $ is not very important. One can see from Eq. (\ref%
{k3}) that the originally sharp dependence of the propagator $D$ on the
momentum $\mathbf{Q-k}$ is smeared due to the random shapes of the domains.
The function $\tilde{U}\left( \left\vert \mathbf{q-Q}\right\vert
/Q_{D}\right) $ should describe a smeared shape of paramagnon peaks in
neutron scattering. Experimentally observed peaks are indeed rather broad
\cite{dai,fong,li}.

The structure of the DW containing both magnetic moments and holes should
result in a coupling of the mode $\bar{D}^{R}$ not only to spin but also to
charge excitations.

\section{Factorization of the imaginary part of self-energy into energy- and
momentum-dependent parts.}

Many physical quantities can be obtained using the imaginary part \textrm{Im}%
$\Sigma ^{R}$ of the self-energy $\Sigma ^{R}$ of the retarded one-particle
electron Green function. A very important feature of the MFL hypothesis is
that \textrm{Im}$\Sigma ^{R}$ factorizes into energy-and momentum dependent
parts \cite{varma1,abrahams0,abrahams}. It is this property that leads
finally the universal dependencies of physical quantities on temperature,
energy, etc.

We calculate \textrm{Im}$\Sigma ^{R}$ using a self-consistent Born
approximation. A standard representation for \textrm{Im}$\Sigma ^{R}\mathrm{%
\ }$reads%
\begin{eqnarray}
&&\mathrm{Im}\Sigma ^{R}\left( \varepsilon ,\mathbf{p}\right) =-\frac{%
\lambda ^{2}}{\left( 2\pi \right) ^{3}}\int d\mathbf{p}_{1}\int_{-\infty
}^{\infty }d\omega \mathrm{Im}G^{R}\left( \varepsilon -\omega ,\mathbf{p}%
_{1}\right)  \notag \\
&&\times \mathrm{Im}\bar{D}^{R}\left( \omega ,\mathbf{p-p}_{1}\right) \left(
\tanh \frac{\varepsilon -\omega }{2T}+\coth \frac{\omega }{2T}\right) ,
\label{d1}
\end{eqnarray}%
where
\begin{equation}
G^{R}\left( \varepsilon ,\mathbf{p}\right) =\left( \varepsilon -\varepsilon
\left( \mathbf{p}\right) +\mu +i/\left( 2\tau \left( \mathbf{p}\right)
\right) \right) ^{-1},  \label{d1a}
\end{equation}%
\begin{equation}
\frac{1}{2\tau \left( \mathbf{p}\right) }=\frac{1}{2\tau _{el}}-\mathrm{Im}%
\Sigma ^{R}\left( \varepsilon ,\mathbf{p}\right)  \label{d1b}
\end{equation}%
and $\tau _{el}$ is the elastic scattering time due to scattering on
non-magnetic impurities. In principle, Eq. (\ref{d1}-\ref{d1b}) is an
integral equation. However, it can easily be solved assuming that the
dependence of $G^{R}\left( \varepsilon ,\mathbf{p}_{1}\right) $ on the
component $p_{1\perp }$perpendicular to the Fermi surface is more sharp than
that of $\mathrm{Im}\bar{D}^{R}\left( \omega ,\mathbf{p-p}_{1}\right) $.
Then, we neglect $p_{1\perp }$ in $\mathrm{Im}\bar{D}^{R}\left( \omega ,%
\mathbf{p-p}_{1}\right) $ and integrate over this variable. The main
contribution comes from the vicinity of the Fermi surface and we obtain
\begin{equation}
\mathrm{Im}\Sigma ^{R}\left( \varepsilon ,\mathbf{p}\right) =-\frac{\lambda
^{2}A\left( \mathbf{p}\right) T}{\left( 4\pi \right) ^{2}}f\left( \frac{%
\varepsilon }{2T}\right) ,  \label{d3}
\end{equation}%
where

\begin{equation}
A\left( \mathbf{p}\right) =Q_{D}^{-2}\int_{FS}\tilde{U}\left( \left\vert
\mathbf{p-\mathbf{Q}-\bar{p}}_{1}\right\vert /Q_{D}\right) \frac{d\mathbf{%
\bar{p}}_{1}}{v\left( \mathbf{\bar{p}}_{1}\right) },  \label{d4}
\end{equation}%
$v\left( \mathbf{\bar{p}}_{1}\right) $ is the velocity at a point $\mathbf{%
\bar{p}}_{1}$on the Fermi surface, the integration is performed over the
Fermi surface, and
\begin{equation}
f\left( u\right) =\int_{-\infty }^{\infty }\left( \tanh \left( u-x\right)
+\coth x\right) \arctan \left( bx\right) dx,  \label{d6}
\end{equation}%
where $b=2\gamma /\left( cg\right) $ is of order $1.$The function $\mathrm{Im%
}\Sigma ^{R}\left( \varepsilon ,\mathbf{p}\right) $ is a smooth function of
the position on the Fermi surface and does not depend on the elastic
scattering time $\tau _{el}$.

The approximation used for the derivation of $\mathrm{Im}\Sigma ^{R}\left(
\varepsilon ,\mathbf{p}\right) $ is applicable for $\tau ^{-1}\left( \mathbf{%
p}\right) \ll v_{F}Q_{D}$, where $v_{F}$ is a typical Fermi velocity. For a
weak scattering on impurities, one comes using Eqs. (\ref{d3},\ref{d4}) to
inequality
\begin{equation}
T\ll T_{1}=\left( Q_{D}v_{F}\right) ^{2}/\lambda ^{2}  \label{d6a}
\end{equation}%
At the same time, the temperature $T^{\ast }$ separating the pseudogap phase
and metallic region was evaluated within the spin-fermion model in Ref. %
\onlinecite{emp} as $T^{\ast }\sim 0.1\lambda ^{2}$, which allows one to
estimate the energy $\lambda ^{2}$ as
\begin{equation}
\lambda ^{2}\sim 2000-3000K  \label{d6b}
\end{equation}%
As $Q_{D}\sim pQ$, we can estimate the energy $Q_{D}v_{F}$ as
\begin{equation}
Q_{D}v_{F}\sim 1000K  \label{d6aa}
\end{equation}%
Using the estimates (\ref{d6a}, \ref{d6aa}) one can conclude that at
temperatures%
\begin{equation}
T\lesssim 300-500K  \label{d6c}
\end{equation}%
the approximation used is clearly justified. Of course, the estimate does
not exclude the linear temperature dependence of $\mathrm{Im}\Sigma
^{R}\left( \varepsilon ,\mathbf{p}\right) $ even at higher temperatures.

It is relevant to mention that the mean free path $l=v_{F}\tau $ may
considerably exceed the domain size $Q_{D}^{-1}.$ Although the domain
borders contain charges, the picture can be smeared due to overlap of the
boarders near the quantum critical point. In addition, the charges can be
screened. All this can reduce the scattering amplitudes and result in a long
elastic mean free path and a rough estimation leads to a conclusion that the
temperature $T_{1}$ can reach values of order $1000K$.

Remarkably, the function $\mathrm{Im}\Sigma ^{R}\left( \varepsilon ,\mathbf{p%
}\right) $, Eq. (\ref{d3}), factorizes into the energy- and
momentum-dependent parts. Therefore, its temperature and energy dependence
is the same for all parts of the Fermi surface. One can write
\begin{equation}
\mathrm{Im}\Sigma ^{R}\left( \varepsilon ,\mathbf{p}\right) \propto -\lambda
^{2}\max \left( \left\vert \varepsilon \right\vert ,T\right)  \label{d8}
\end{equation}%
in agreement with the findings of Refs. %
\onlinecite{varma1,abrahams0,abrahams}.

The electron spectral function has been compared in Ref. %
\onlinecite{abrahams} with the results of the ARPES measurements of Refs. %
\onlinecite{kaminski,valla} and a good agreement has been found. Using Eq. (%
\ref{k3}) one can describe also the other experiments discussed in Refs. %
\onlinecite{varma1,abrahams0,abrahams} and, in particular, obtain linear in
temperature d.c. resistivity.

\section{Linear temperature dependence of resistivity.}

Having calculated the imaginary part \textrm{Im}$\Sigma ^{R}\left(
\varepsilon ,\mathbf{p}\right) \mathrm{\ }$of the self-energy $\Sigma
^{R}\left( \varepsilon ,\mathbf{p}\right) ,$ Eqs. (\ref{d4}-\ref{d6}), we
can calculate the conductivity and resistivity. The zero frequency
conductivity $\sigma $ can conveniently be calculated using the
Kubo-Kirkwood formula%
\begin{equation}
\sigma =\frac{2e^{2}}{\pi }\int v_{x}^{2}\left( \mathbf{p}\right) \left[
\text{\textrm{Im}}G^{R}\left( \mathbf{p}\right) \right] ^{2}\frac{d\mathbf{p}%
}{\left( 2\pi \right) ^{2}},  \label{m1}
\end{equation}%
where $v_{x}\left( \mathbf{p}\right) $ is the $x$-component of the velocity,
$G^{R}$ is the retarded Green function taken at zero energy $\varepsilon $
and averaged over all types of disorder.

In principle, the integrand in Eq. (\ref{m1}) should contain the disorder
average of the product of the Green functions. However, neglecting
localization effects this fact is important only in the case of a smooth
disorder. In the latter case one should simply replace at the end the
scattering time $\tau $ by a longer transport time $\tau _{tr}$. As we
consider scattering with the large vector $\mathbf{Q},$ just writing
averaged Green functions can be a good approximation.

We write the Green function $G^{R}\left( \mathbf{p}\right) $ as
\begin{equation}
G^{R}\left( \mathbf{p}\right) =-\left[ \varepsilon \left( \mathbf{p}\right)
-\mu -\frac{i}{2\tau _{el}}+i\text{\textrm{Im}}\Sigma ^{R}\left( \mathbf{p}%
\right) \right] ^{-1},  \label{m2}
\end{equation}%
where $\tau _{el}$ is the elastic scattering time, $\mu $ is the chemical
potential and \textrm{Im}$\Sigma ^{R}\left( \mathbf{p}\right) $ is obtained
from Eq. (\ref{d4}-\ref{d6}) by putting $\varepsilon =0$. We write this
function as
\begin{equation}
\text{\textrm{Im}}\Sigma ^{R}\left( \mathbf{p}\right) =-\frac{\lambda
^{2}A\left( \mathbf{p}\right) Tf\left( 0\right) }{\left( 4\pi \right) ^{2}},
\label{m3}
\end{equation}%
where the function $A\left( \mathbf{p}\right) $ and $f\left( 0\right) $ are
determined by Eqs. (\ref{d4}-\ref{d6}).

This allows one to express the conductivity $\sigma $ in terms of the
following integral%
\begin{equation}
\sigma =\frac{e^{2}}{\pi }\int \frac{\mathbf{v}^{2}\left( \mathbf{p}\right)
}{\left[ \left( \varepsilon \left( \mathbf{p}\right) -\mu \right) ^{2}+\frac{%
1}{4\tau ^{2}\left( \mathbf{p}\right) }\right] ^{2}}\frac{1}{4\tau
^{2}\left( \mathbf{p}\right) }\frac{d\mathbf{p}}{\left( 2\pi \right) ^{2}},
\label{m4}
\end{equation}%
where $\tau \left( \mathbf{p}\right) $ equals
\begin{equation}
\frac{1}{2\tau \left( \mathbf{p}\right) }=\frac{1}{2\tau _{el}}+\frac{%
\lambda ^{2}A\left( \mathbf{p}\right) Tf\left( 0\right) }{\left( 4\pi
\right) ^{2}}  \label{m6}
\end{equation}

As $\tau ^{-1}\left( \mathbf{p}\right) $ is assumed to be not very large,
such that the inequality
\begin{equation}
\tau ^{-1}\left( \mathbf{p}\right) \ll Q_{D}v_{F}  \label{m7}
\end{equation}
is fulfilled, the main contribution into the integral (\ref{m4}) comes from
the narrow region near the Fermi surface. This allows one to integrate
separately over the perpendicular to the Fermi surface component $\mathbf{p}%
_{\perp }$ of the momentum using the variable $\xi =\varepsilon \left(
\mathbf{p}\right) -\mu \simeq \left( \mathbf{p}_{\perp }-\mathbf{\bar{p}}%
\right) \mathbf{v}\left( \mathbf{\bar{p}}\right) $ and the vector on the
Fermi surface $\mathbf{\bar{p}.}$

Integrating over $\xi $ we reduce the conductivity $\sigma $ to the form%
\begin{equation}
\sigma =e^{2}\int_{FS}v\left( \mathbf{\bar{p}}\right) \tau \left( \mathbf{%
\bar{p}}\right) \frac{d\mathbf{\bar{p}}}{\left( 2\pi \right) ^{2}}\mathbf{,}
\label{m8}
\end{equation}%
where $v\left( \mathbf{\bar{p}}\right) =\left\vert \mathbf{v}\left( \mathbf{%
\bar{p}}\right) \right\vert $ is the modulus of the velocity at the momentum
$\mathbf{\bar{p}}$ on the Fermi surface and the integration in Eq. (\ref{m8}%
) is performed over the Fermi surface.

Actually, we assume that the main contribution to $\tau ^{-1}\left( \mathbf{p%
}\right) ,$ Eq. (\ref{m6}), comes from $\mathrm{Im}\Sigma ^{R}\left(
\varepsilon ,\mathbf{p}\right) $ and the inequality (\ref{d6a}) is fulfilled.

Using Eq. (\ref{m6}) we write the resistivity $\rho $ as
\begin{equation}
\rho \left( T\right) =\frac{1}{e^{2}\nu _{eff}}\left[ \left\langle \frac{%
v^{2}\left( \mathbf{\bar{p}}\right) }{2}\left( \frac{1}{2\tau _{el}}+\frac{%
\lambda ^{2}A\left( \mathbf{\bar{p}}\right) Tf\left( 0\right) }{\left( 4\pi
\right) ^{2}}\right) ^{-1}\right\rangle _{FS}\right] ^{-1}  \label{m9}
\end{equation}%
where the symbol $\left\langle ...\right\rangle _{FS}$ stands for the
average over the Fermi surface%
\begin{equation}
\left\langle ...\right\rangle _{FS}=\frac{1}{\nu _{eff}}\int_{FS}\frac{%
\left( ...\right) }{v\left( \mathbf{\bar{p}}\right) }\frac{d\mathbf{\bar{p}}%
}{\left( 2\pi \right) ^{2}},  \label{m10}
\end{equation}%
and $\nu _{eff}$ is given by the integral
\begin{equation}
\nu _{eff}=\frac{1}{\left( 2\pi \right) ^{2}}\int_{FS}\frac{d\mathbf{\bar{p}}%
}{v\left( \mathbf{\bar{p}}\right) }  \label{m11}
\end{equation}%
The quantity $\nu _{eff}$ is the standard density of states per spin
direction for a circular Fermi surface but it may numerically differ from
the latter for more complex geometries.

In case of a large $Q_{D}$, when the domain size is of the same order as
atomic distances or slightly exceeds the latter, the function $A\left(
\mathbf{\bar{p}}\right) $ weakly depends on the momenta $\mathbf{\bar{p}}$
on the Fermi surface. This possibility is supported by the fact that there
are $8$ hot spots in the Brillouin zone and the distance between them may be
somewhat smaller than the antiferromagnetic vector $\mathbf{Q}$. If one
neglected the dependence of $A\left( \mathbf{\bar{p}}\right) $ on $\mathbf{%
\bar{p}}$ one would obtain the resistivity $\rho \left( T\right) $ simply
putting in Eq. (\ref{m9}) $A\left( \mathbf{\bar{p}}\right) =A.$ In this
case, the averaging over the Fermi surface in Eq. (\ref{m9}) is trivial and
the resistivity $\rho \left( T\right) $ takes the form%
\begin{equation}
\rho \left( T\right) =\rho _{0}+\alpha T,  \label{m12}
\end{equation}%
where $\rho _{0}$ is the residual resistivity and
\begin{equation}
\alpha =\frac{\lambda ^{2}Af\left( 0\right) }{\left( 4\pi e\right) ^{2}E}%
,\quad E=\frac{1}{2}\int_{FS}v\left( \mathbf{\bar{p}}\right) \frac{d\mathbf{%
\bar{p}}}{\left( 2\pi \right) ^{2}}.  \label{m13}
\end{equation}%
In Eq. (\ref{m13}) the parameter $E$ is an energy of the order of the Fermi
energy.

The ratio of the first and second term in Eq. (\ref{m12}) can be arbitrary
and, in particular, the $T$-dependent term can be much larger that the
residual resistivity $\rho _{0}.$

As concerns the lower limit, the temperature $T$ should not be in the
pseudogap region, which gives the inequality%
\begin{equation}
T>T^{\ast }  \label{m14}
\end{equation}

In reality, at finite $\tau _{el}$ the resistivity $\rho \left( T\right) $
is not universally linear in $T$ due to a dependence of $A\left( \mathbf{%
\bar{p}}\right) $ on the momentum $\mathbf{\bar{p}}$ on the Fermi surface.
Nevertheless, it does become linear at sufficiently high temperatures. This
can be seen from the expansion in small $\left( \tau _{el}T\right) ^{-1}$ of
the resistivity $\rho \left( T\right) $ in Eq. (\ref{m9}). The calculation
is straightforward and one can easily write the first three terms of the
expansion of $\rho \left( T\right) $%
\begin{equation}
\rho \left( T\right) =\left( e^{2}\nu _{eff}\right) ^{-1}\left( b_{1}T+\frac{%
b_{0}}{\tau _{el}}+\frac{b_{-1}}{\tau _{el}^{2}T}\right) ,  \label{m15}
\end{equation}%
\begin{equation}
b_{1}=\frac{\lambda ^{2}f\left( 0\right) }{8\pi ^{2}}\left\langle \frac{%
v^{2}\left( \mathbf{\bar{p}}\right) }{A\left( \mathbf{\bar{p}}\right) }%
\right\rangle _{FS}^{-1},  \label{m16}
\end{equation}%
\begin{equation}
b_{0}=\left\langle \frac{v^{2}\left( \mathbf{\bar{p}}\right) }{A^{2}\left(
\mathbf{\bar{p}}\right) }\right\rangle _{FS}\left\langle \frac{v^{2}\left(
\mathbf{\bar{p}}\right) }{A\left( \mathbf{\bar{p}}\right) }\right\rangle
_{FS}^{-2},  \label{m17}
\end{equation}%
\begin{eqnarray}
&&b_{-1}=\frac{8\pi ^{2}}{\lambda ^{2}f\left( 0\right) }\left\langle \frac{%
v^{2}\left( \mathbf{\bar{p}}\right) }{A\left( \mathbf{\bar{p}}\right) }%
\right\rangle _{FS}^{-3}  \label{m18} \\
&&\times \Big[\left\langle \frac{v^{2}\left( \mathbf{\bar{p}}\right) }{%
A^{2}\left( \mathbf{\bar{p}}\right) }\right\rangle _{FS}^{2}-\left\langle
\frac{v^{2}\left( \mathbf{\bar{p}}\right) }{A\left( \mathbf{\bar{p}}\right) }%
\right\rangle _{FS}\left\langle \frac{v^{2}\left( \mathbf{\bar{p}}\right) }{%
A^{3}\left( \mathbf{\bar{p}}\right) }\right\rangle _{FS}\Big]  \notag
\end{eqnarray}%
It is clear from Eqs. (\ref{m15}-\ref{m18}) that the coefficient $b_{-1}$ in
the third term in Eq. (\ref{m15}), as well as all higher terms of the
expansion in $\left( \tau _{el}T\right) ^{-1}$, vanishes in the case when $%
A\left( \mathbf{\bar{p}}\right) $ does not depend on the momentum $\mathbf{%
\bar{p}}$ on the Fermi surface and one comes to Eq. (\ref{m12}). If $A\left(
\mathbf{\bar{p}}\right) $ depends on the $\mathbf{\bar{p}}$ the third term
in Eq. (\ref{m15}) is finite but it is small in the limit $\tau _{el}T\gg 1$%
. The characteristic temperature $T_{1}$ of the deviation from the linear
dependence depends on the form of the function $A\left( \mathbf{\bar{p}}%
\right) $. One can roughly estimate this temperature as
\begin{equation}
T_{0}=\frac{\left( 4\pi \right) ^{2}\left( A_{\text{\textrm{min}}}^{-1}-A_{%
\text{\textrm{max}}}^{-1}\right) }{2\tau _{el}\lambda ^{2}f\left( 0\right) },
\label{m19}
\end{equation}%
where $A_{\text{\textrm{max}}}$ and $A_{\text{\textrm{min}}}$ are maximum
and minimum values of $A\left( \mathbf{\bar{p}}\right) $ on the Fermi
surface. One obtains the linear in $T$ behavior for temperatures $T\gtrsim
T_{0}$. Of course, the inequality (\ref{m14}) should also be fulfilled.

Thus, using Eqs. (\ref{m15}, \ref{m19}) we come to the conclusion that the
region of the linear resistivity exists provided the following inequality is
fulfilled%
\begin{equation}
\tau _{el}^{-1}\ll \left( Q_{D}v_{F}\right) ^{2}\frac{A_{\text{\textrm{max}}%
}A_{\text{\textrm{min}}}}{A_{\text{\textrm{max}}}-A_{\text{\textrm{min}}}}
\label{m20}
\end{equation}%
Estimating typical values of $A$ as $A_{\text{\textrm{max}}}\sim \left(
v_{F}Q_{D}\right) ^{-1}$ and introducing a parameter $\kappa =A_{\text{%
\textrm{max}}}/A_{\text{\textrm{min}}}$ we rewrite the inequality (\ref{m20}%
) as
\begin{equation}
\tau _{el}^{-1}\ll \frac{Q_{D}v_{F}}{\kappa -1}  \label{m21}
\end{equation}

Of course, a linear temperature dependence can also be obtained in the limit
$T\ll T_{0}$ when the main contribution to the resistivity comes from the
scattering on impurities. However, this limit is not as interesting as the
opposite limit of high temperatures.

\section{Discussion and comparison with experimental data.}

A model of fermions interacting with antiferromagnetic spin fluctuations has
been considered. It was assumed that the interaction is random due to
presence of domains with different phase of the antiferromagnetic field. As
a microscopic mechanism supporting existence of these \textquotedblleft $\pi
$-shifted domains\textquotedblright , it was assumed that stripes are formed
in the doped antiferromagnet and eventually destroy the antiferromagnetic
order affecting, however, antiferromagnetic fluctuations in the metallic
side. This is a new type of disorder that has not previously been considered
in models of cuprates. Of course, conventional potential disorder can be
present in the model as well.

Although the model is quite simple, it allows one to obtain results
predicted on basis of the MFL hypothesis \cite{varma1,abrahams0} in a rather
simple way. Of course, the model considered here is not free of assumptions
and is not completely microscopical. However, it is definitely
\textquotedblleft more microscopic\textquotedblright\ than the MFL
hypothesis of Refs. \onlinecite{varma1,abrahams0}. It has also predictive
power being able to describe the dependence of the slope of the linear
temperature dependence of the resistivity on doping.

The slope $\alpha $ of the $T$-dependence does not depend on $\tau _{el}$
but the residual resistivity $\rho _{0}$ determined by $\tau _{el}$ does.
This agrees with observations of Ref. \onlinecite{albenque}. At the same
time, a clear decrease of the slope with the doping has been observed in
experiments \onlinecite{ando,cooper}. A more detailed microscopic theory is
necessary in order to describe precisely the dependence of $\alpha $ on the
doping $p$ but a rough estimation can already be done using Eqs. (\ref{d4}, %
\ref{m12}, \ref{m13}).

\begin{figure}[h]
\vspace{-0.5cm} \includegraphics[height=7in,width=1.2\linewidth]{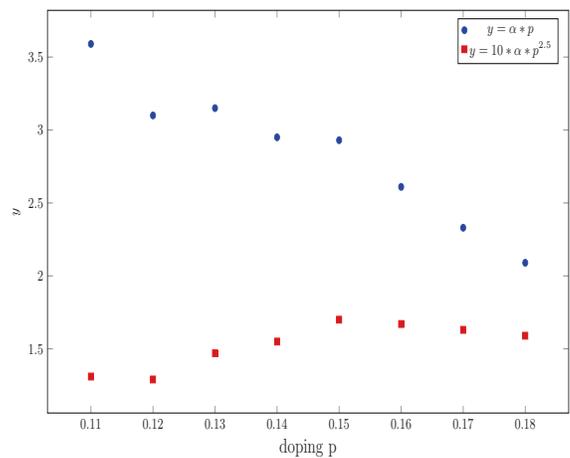}
\center
\vspace{-11.0cm}
\caption{(Color online.) Dependence of $y=\protect\alpha p$ (blue dots) and $%
y=10\protect\alpha p^{2.5}$ (red boxes) on doping $p$ extracted from Fig.1a
of Ref. \onlinecite{ando}}
\label{fig02}
\end{figure}
We use for comparison between theory and experiment Fig.1a of Ref. %
\onlinecite{ando} displaying the linear temperature dependence of
resistivity Bi$_{2}$Sr$_{2-x}$La$_{x}$CuO$_{6+\delta }$ for doping $p$ $%
=0.11-0.18$. The slope $\alpha $ is extracted from the difference $\Delta
\rho =\rho \left( 300K\right) -\rho \left( 100K\right) .$ Estimating
physical quantities characterizing the fermions we simply assume that their
density (volume under Fermi surface) is proportional to $pQ^{2}$ and $%
Q_{D}\sim pQ$ ($Q$ is inverse interatomic atomic distance). It also assumed
that there are no singularities on the Fermi surface.

It is important to emphasize that the spin-fermion (SF) model contains low
energy effective fermions instead of original electrons on the $CuO$
lattice. The shape of the Fermi surface of these fermions and the dependence
of the Fermi energy on doping is formally not specified in the SF model and
one is to be guided by reasonable assumptions. As the SF model is designed
to describe the system near QCP, one has no need to think on what happens in
the limit $p\rightarrow 0$ when the system becomes a Mott insulator. At the
same time, in the vicinity of QCP one can reasonably assume that the density
of the fermions in SF model is proportional to the density of doped
electron, which leads to the proportionality of the fermion density to $%
pQ^{2}$. This proportionality is clearly good for comparatively high doping.
As concerns low doping, it may still be a good approximation in the
framework of SF model even in the antiferromagnetic region provided one
stays in the vicinity of QCP.

Using the original formulation of MFL, Eq. (\ref{a1}), of Refs. %
\onlinecite{varma1,abrahams0} and the fact that in SF model the density of
states $\nu $ is thus independent in 2D of doping $p$ one comes to the
relation $\alpha \varpropto v_{F}^{-2}\varpropto p^{-1}$. The dependence of $%
y=\alpha p$ on $p$ taken from Fig.1a of Ref. \onlinecite{ando} is
represented by dots in Fig. \ref{fig02}. Its essential dependence on $p$
indicates that Eq. (\ref{a1}) should possibly be modified. At the same time,
it follows from Eq. (\ref{d4}) that $A\varpropto \left( v_{F}Q_{D}\right)
^{-1}$ and $E\varpropto mv_{F}^{2}$, which leads to $\alpha \varpropto
v_{F}^{-3}Q_{D}^{-1}\varpropto p^{-5/2}$. The variation of $y=10\alpha
p^{5/2}$ with $p$ is represented by boxes in Fig. \ref{fig02}. A weak
dependence of $y=\alpha p^{5/2}$ on the doping $p$ supports the present
theory. As the discussion presented here is based on the assumption that the
doping is not too low, it is important to emphasize that the lowest doping
level $p$ studied in Ref. \onlinecite{ando} is $p=0.11,$ which is already
well in the metallic region. Therefore, the assumption that the density of
states is weakly dependent on doping is not unrealistic for $p\geq 0.11$.

Anyway, the quantity $y=10\alpha p^{5/2}$ in Fig. \ref{fig02} is not exactly
a constant and one can speak rather of a qualitative agreement than of a
microscopic theory. However, the present formulation already gives a better
agreement with the experimental data than the original version of the MFL
hypothesis, Eq. (\ref{a1}). Actually, to the best of our knowledge, the
dependence of the slope on the doping is discussed here for the first time
and the theory presented has a potential of further improvement.

In conclusion, fermions interacting with critical antiferromagnetic
fluctuations in two dimensions are considered. Assuming that the $CuO$
planes consist of different domains, such that the coupling constant $%
\lambda $ changes the sign when crossing the boarders between them, we have
derived the hypothetical mode of the Marginal Fermi Liquid and clarified its
dependence on the doping. The slope of the linear temperature dependence of
the resistivity calculated here is compared with experimental results and an
encouraging agreement is found.

\begin{acknowledgments}
The author gratefully acknowledges the financial support of the Ministry of
Education and Science of the Russian Federation in the framework of Increase
Competitiveness Program of NUTS~\textquotedblleft MISiS\textquotedblright\
(Nr.~K2-2014-015) and of Transregio 12 of \textit{Deutsche
Forschungsgemeinschaft}.
\end{acknowledgments}

\appendix

\section{Calculation of $\mathrm{Im}\bar{D}\left( \protect\omega ,\mathbf{q}%
,T\right) $.}

Here we calculate the function $\mathrm{Im}\bar{D}\left( \omega ,\mathbf{q}%
,T\right) $, where $\bar{D}^{R}\left( \omega ,\mathbf{q,}T\right) $ is the
analytical continuation $i\omega _{n}\rightarrow \omega +i\delta $ from
Matsubara frequencies $\omega _{n}$ to real frequencies $\omega $ of the
function $\bar{D}\left( \omega _{n},\mathbf{q}\right) $, Eq. (\ref{k2})
\begin{equation}
\bar{D}\left( \omega _{n},\mathbf{q}\right) =Q_{D}^{-2}\int \tilde{U}\left(
\frac{\left\vert \mathbf{q-k}\right\vert }{Q_{D}}\right) D\left( \omega _{n},%
\mathbf{k}\right) \frac{d\mathbf{k}}{\left( 2\pi \right) ^{2}},  \label{ap1}
\end{equation}%
with $D\left( \omega _{n},\mathbf{k}\right) $ from Eq. (\ref{c11}) of the
main text.
\begin{equation}
D\left( \omega _{n},\mathbf{k}\right) =\left( \gamma \left\vert \omega
_{n}\right\vert +\left( \mathbf{Q-k}\right) ^{2}+a\left( T\right) \right)
^{-1}.  \label{ap2}
\end{equation}%
The analytical continuation of propagator $D\left( \omega _{n},\mathbf{k}%
\right) $ can easily be performed leading to the retarded propagator%
\begin{equation}
D^{R}\left( \omega ,\mathbf{k}\right) =\left( -i\gamma \omega +\left(
\mathbf{Q-k}\right) ^{2}+a\left( T\right) \right) ^{-1}.  \label{ap3}
\end{equation}%
Then, we obtain for the imaginary part of this function the following
expression%
\begin{equation}
\mathrm{Im}D^{R}\left( \omega ,\mathbf{k}\right) =\frac{\gamma \omega }{%
\gamma ^{2}\omega ^{2}+\left( \left( \mathbf{Q-k}\right) ^{2}+a\left(
T\right) \right) ^{2}}  \label{ap4}
\end{equation}%
Using Eq. (\ref{ap4}) we represent $\mathrm{Im}\bar{D}\left( \omega ,\mathbf{%
q,}T\right) $ in the form%
\begin{eqnarray}
&&\text{\textrm{Im}}\bar{D}\left( \omega ,\mathbf{q,}T\right)  \label{ap5} \\
&=&Q_{D}^{-2}\int \frac{\gamma \omega \tilde{U}\left( \frac{\left\vert
\mathbf{q-k}\right\vert }{Q_{D}}\right) }{\gamma ^{2}\omega ^{2}+\left(
\left( \mathbf{Q-k}\right) ^{2}+a\left( T\right) \right) ^{2}}\frac{d\mathbf{%
k}}{\left( 2\pi \right) ^{2}}  \notag
\end{eqnarray}%
Shifting in the integral the momentum $\mathbf{k\rightarrow k+Q}$ the
function $\mathrm{Im}\bar{D}\left( \omega ,\mathbf{q,}T\right) $ can be
written as

\begin{eqnarray}
&&\text{\textrm{Im}}\bar{D}\left( \omega ,\mathbf{q,}T\right)  \label{ap6} \\
&=&Q_{D}^{-2}\int_{0}^{\infty }\int_{0}^{2\pi }\frac{\gamma \omega \tilde{U}%
\left( \frac{\sqrt{k^{2}-2k\left\vert \mathbf{q-Q}\right\vert \cos \theta
+\left\vert \mathbf{q-Q}\right\vert ^{2}}}{Q_{D}}\right) }{\gamma ^{2}\omega
^{2}+\left( k^{2}+a\left( T\right) \right) ^{2}}\frac{kdkd\theta }{\left(
2\pi \right) ^{2}},  \notag
\end{eqnarray}%
where $\theta $ is the angle between the vectors $\mathbf{k}$ and $\mathbf{%
q-Q}$.

In the limit $\gamma \left\vert \omega \right\vert \ll Q_{D}^{2},$ the main
contribution to the integral in Eq. (\ref{ap6}) comes from $k\sim \left(
\gamma \omega \right) ^{1/2}\ll Q_{D}$. This allows one to neglect $k$ in
the argument of the function $\tilde{U}$. Changing the variable of the
integration to $z=k^{2}$ we come to the integral

\begin{equation}
\text{\textrm{Im}}\bar{D}\left( \omega ,\mathbf{q,}T\right)
=Q_{D}^{-2}\int_{0}^{\infty }\frac{\gamma \omega \tilde{U}\left( \left\vert
\mathbf{q-Q}\right\vert /Q_{D}\right) }{\gamma ^{2}\omega ^{2}+\left(
z+a\left( T\right) \right) ^{2}}\frac{dz}{4\pi },  \label{ap7}
\end{equation}%
Calculating the integral over $z$ we come to Eq. (\ref{k3}) of the main text.

\section{Calculation of $\mathrm{Im}\Sigma ^{R}$.}

A convenient representation of the imaginary part \textrm{Im}$\Sigma
^{R}\left( \varepsilon ,\mathbf{p}\right) $ can be found in Eqs. (\ref{d1}-%
\ref{d1b}) of the main text and we use it here. Eqs. (\ref{d1}-\ref{d1b})
are written in the self- consistent Born approximation. It can be obtained
writing the Green functions on Matsubara frequencies and making analytical
continuation to frequencies $\omega $ on the real axis. As the Green
function in the integrand contains $\mathrm{Im}\Sigma ^{R}\left( \varepsilon
,\mathbf{p}\right) $, Eq. (\ref{d1}) is an integral equation and one should
solve this equation in order to find this quantity. The solution is rather
simple in the case when the dependence of imaginary part $\mathrm{Im}%
G^{R}\left( \varepsilon ,\mathbf{p}_{1}\right) $ of the Green function $%
G^{R} $ on the component $\mathbf{p}_{1\perp }$ is more sharp than the
dependence of $\mathrm{Im}\bar{D}^{R}\left( \omega ,\mathbf{p-p}_{1}\right) $
on the same variable. In this situation, one may simply replace $\mathbf{p}%
_{1}$ in $\mathrm{Im}\bar{D}^{R}\left( \omega ,\mathbf{p-p}_{1}\right) $ by
its value $\mathbf{\bar{p}}_{1}$ on the Fermi surface and calculate
explicitly the integral over $\mathbf{p}_{1\perp }$ in Eq. (\ref{d1}) using
Eq. (\ref{d1a}).

Using Eq. (\ref{d1a}) and integrating $\mathrm{Im}G^{R}\left( \varepsilon
-\omega ,\mathbf{p}_{1}\right) $ over $\mathbf{p}_{1\perp }$ while keeping
the parallel component of $\mathbf{p}$ fixed at a point $\mathbf{\bar{p}}%
_{1} $ on the Fermi surface we have%
\begin{equation}
\int \mathrm{Im}G^{R}\left( \varepsilon -\omega ,\mathbf{p}_{1}\right) d%
\mathbf{p}_{1\perp }=\int_{-\infty }^{\infty }\mathrm{Im}G^{R}\left(
\varepsilon -\omega ,\mathbf{p}_{1}\right) \frac{d\xi _{1}}{v\left( \mathbf{%
\bar{p}}_{1}\right) },  \label{ap7d}
\end{equation}%
where $\xi _{1}=\varepsilon \left( \mathbf{p}_{1}\right) -\mu \simeq \mathbf{%
v}\left( \mathbf{\bar{p}}_{1}\right) \left( \mathbf{p}_{1}\mathbf{-\bar{p}}%
_{1}\right) $ and $\mathbf{v}\left( \mathbf{\bar{p}}_{1}\right) $ is the
velocity on the Fermi surface at the point $\mathbf{\bar{p}}_{1}$.
Neglecting the perpendicular component $\mathbf{p}_{1}-\mathbf{\bar{p}}_{1}$
in $\tau \left( \mathbf{p}_{1}\right) $ we obtain%
\begin{eqnarray}
&&\int \mathrm{Im}G^{R}\left( \varepsilon -\omega ,\mathbf{p}_{1}\right) d%
\mathbf{p}_{1\perp }  \label{ap7e} \\
&=&-\frac{1}{2\tau \left( \mathbf{\bar{p}}_{1}\right) }\int_{-\infty
}^{\infty }\frac{d\xi }{\left( \varepsilon -\xi \right) ^{2}+\left( 2\tau
\left( \mathbf{\bar{p}}_{1}\right) \right) ^{-2}}=\pi  \notag
\end{eqnarray}%
Substituting Eq. (\ref{ap7e}) into Eq. (\ref{d1}) and using Eq. (\ref{k3})
we come immediately to Eqs. (\ref{d3}-\ref{d6}).

\newpage

\end{document}